# Artificial piezoelectricity in centrosymmetric SrTiO$_3$

**B. Khanbabaee**[a*], **E. Mehner**[b], **C. Richter**[b,c], **J. Hanzig**[b], **M. Zschornak**[b], **U. Pietsch**[a], **H. Stöcker**[b], **T. Leisegang**[b], **D. C. Meyer**[b], **and S. Gorfman**[a*]

[a] Department of Physics, University of Siegen, 57072 Siegen, Germany

[b] Institute of Experimental Physics, TU Bergakademie Freiberg, 09596 Freiberg, Germany

[c] Deutsches Elektronen-Synchrotron / DESY - Photon Science, 22607 Hamburg, Germany

\* khanbabaee@physik.uni-siegen.de

\* gorfman@physik.uni-siegen.de

**Abstract**

Defect engineering is an effective and powerful tool to control existing material properties and create completely new ones, which are symmetry-forbidden in a defect-free crystal. This letter reports on the creation of piezoelectrically active near-surface layer of centrosymmetric SrTiO$_3$, modified by the electric field-induced migration of oxygen vacancies. We provide the unequivocal proof of piezoelectricity through the stroboscopic time-resolved X-ray diffraction under alternating electric field. The magnitude of the discovered piezoelectric effect is comparable with the bulk piezoelectric effect in commercial ferroelectric materials. Such artificially formed defect-mediated piezoelectricity can be important as an alternative road for smart materials design.

Transition-metal oxides attract strong interest due to the large variety of applications [1-6]. Their properties are adjustable through defect engineering: for example, the variation of oxygen partial pressure can alter the electrical conductivity of a solid from $p$– to $n$–type [7] or even change the material from an insulator into a conductor [8]. The properties can be also influenced by the migration of oxygen vacancies [9-17]: it is known that the application of a static external electric field to a room temperature SrTiO$_3$ (STO) single crystal forms the near-surface migration-induced field-stabilized polar (MFP) phase [11]. The physical properties of this phase strongly differ from that of the cubic ($Pm\bar{3}m$ space group) bulk STO. The MFP phase exhibits pyroelectricity, which is symmetry-forbidden in the bulk STO. It suggests that the MFP phase belongs to a non-centrosymmetric subgroup of $Pm\bar{3}m$, e.g. tetragonal *P4mm* [18]. Such symmetry lowering may lead to the appearance of piezoelectricity, which is also forbidden in all centrosymmetric materials.

Despite the great technological importance of piezoelectricity, there is only a very limited stock of commonly available piezoelectric materials [19]. The urgent need to develop environmentally friendly alternatives to the market dominating PbZr$_{1-x}$Ti$_x$O$_3$ (PZT) compounds motivates for the active investigation of structural reasons of piezoelectricity. Alternatively, defect engineering could be considered as a route to design new piezoelectrics. Surprisingly, such artificially created piezoelectricity deserved very little attention so far.

This letter reports the observation of piezoelectricity in the room-temperature MFP phase of STO, through the stroboscopic *in-situ* time-resolved X-ray diffraction under alternating electric field. We demonstrate that the piezoelectric coefficients of the MFP phase may even compete with the state-of-the-art materials such as PbZr$_{1-x}$Ti$_x$O$_3$ [21]. The emergence of migration-induced field-stabilized piezoelectricity opens unprecedented opportunities for smart material design and controllable properties.

We used (001) oriented STO single crystal plates (CrysTec, Berlin) of 0.1 mm and 0.5 mm thickness. Titanium electrodes were deposited onto (001) faces to apply a homogenous [001]-directed electric field. The measurements on the 0.1 mm thick sample were performed at an X-ray energy of 8.04 keV using a high resolution home-laboratory (GE HR-XRD 3003) 4-circle diffractometer, Si(111) analyzer crystal and a scintillation detector. The measurements of the 0.5 mm thick sample were performed at the XMaS beamline (BM28, ESRF, Grenoble, France) at an X-ray energy of 16 keV [22]. External voltage was produced by function generator (HAMEG) combined with a high-voltage amplifier (MATSUSADA).

Figure 1 demonstrates the time frames of the experiments with three different voltage regimes. First (Region I in Figure 1), reciprocal space maps (RSMs) were collected at zero electric field. Then (Region II in Figure 1), the MFP phase was formed by the application of a static electric field $E_0 = 1$ kV/mm for 12 hours (as described in [11]). The gradual formation of a "shoulder" at lower values of scattering angles was observed until saturation. Finally (Region III in Figure 1) the dynamics of radial ($\omega$-$2\theta$) scans at the 002 and 101 reflections of the 0.1 mm thick sample, and the 006 reflection of the 0.5 mm thick sample were monitored as a function of time and alternating electric field. Here, a triangular shaped 1 kHz electric field with an amplitude of $\Delta E = 1$ kV/mm was added to the static component $E_0$, so that the electric field varied between $E_0 - \Delta E$ and $E_0 + \Delta E$. While the momentary electric field E(t) may reach zero, the average field was equal to $E_0$. We used a specially designed stroboscopic data-acquisition system operating on the principle of a multichannel analyzer as described in [23-31]. The diffraction signal was distributed between 10000 time-channels, each 100 ns long and the intensities of the different time-channels have been accumulated during 100 s (home lab X-ray diffractometer) and 10 s (ESRF) exposure time per one diffractometer angle.

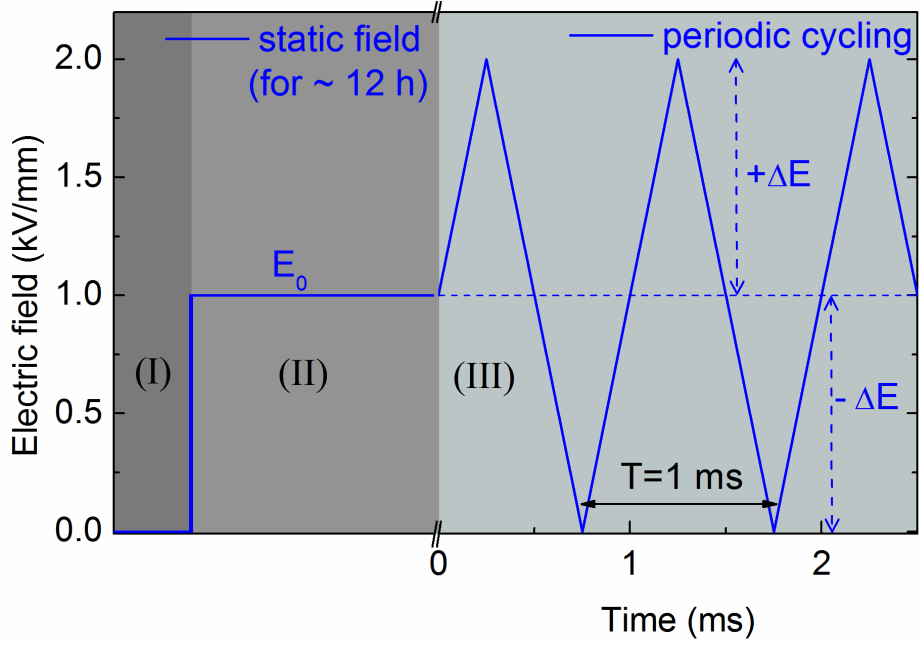

**Figure 1.** Time frames of the experiments. Region I: Probing initial state of the STO. Region II: Monitoring of the formation of the MFP phase under static electric field $E_0 = 1$ kV/mm. Region III: Investigation of the dynamics of the formed MFP phase by the 1 kHz triangular periodic component $\Delta E = \pm 1$ kV/mm added to the static field $E_0$.

Fig. 2 shows RSMs around the 002 and 101 reflections prior to the application of the electric field (a, b) and after the formation of the MFP phase (c, d). The broadening of the peaks perpendicular to the reciprocal lattice vector (along $q_\perp$) characterizes the crystal mosaicity. The appearance of the shoulder peaks parallel to the reciprocal lattice vector (along $q_\parallel$), with lower momentum transfer indicates the formation of the MFP phase with increased lattice spacing. Assuming that the application of electric field induces a tetragonal distortion [11,18], we calculated the corresponding variation of the tetragonal lattice parameters $a$ and $c$ of the 0.1 mm sample (with respect to the bulk STO) as $\Delta c = 5.4(6) \cdot 10^{-3}$ Å and $\Delta a = -0.7(7) \cdot 10^{-3}$ Å. The formed MFP phase of the 0.5 mm sample reveals the larger strain with $\Delta c = 18.8(4) \cdot 10^{-3}$ Å. Therefore, a fivefold increase in sample thickness increased the strain of the MFP phase, relative to the bulk by a factor of 3.4.

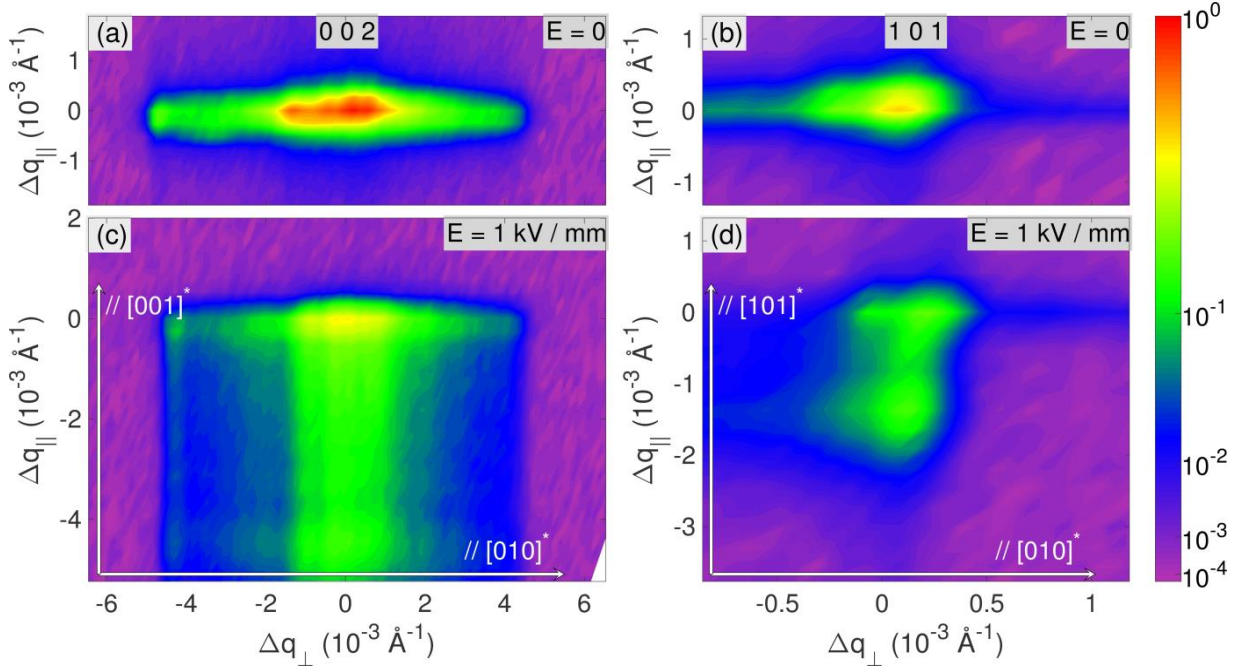

**Figure 2.** Reciprocal space maps of 002 and 101 reflections, collected from the 0.1 mm thick SrTiO$_3$ single crystal before (a-b) and after (c-d) application of a static electric field of $E_0$ = 1 kV/mm for 12 hours. The axes refer to the deviation from the corresponding exact Bragg peak position of bulk STO in the directions, which are parallel ($q_{//}$) or perpendicular ($q_\perp$) to the reciprocal lattice vector. The lower-$q_{//}$ shoulders in figures (c-d) correspond to the diffraction caused by the formed MFP phase.

Figs. 3(a-c) show the false-colour maps of the stroboscopically measured radial (along $q_{//}$) reciprocal space scans as a function of time within the 1 ms period of the added 1 kHz component of electric field. All the intensity maps contain two peaks: the sharp and intense peaks on the right-hand side are diffracted from the bulk STO. The weaker and broader peaks on the left-hand side are diffracted from the near-surface MFP phase. It is clearly seen that the MFP phase peak shifts towards lower $q_{//}$-values as the electric field increases from 0 to 2 kV/mm for t < 0.5 ms and towards higher $q_{//}$-values for t ≥ 0.5 ms.

Figs. 3 (d-f) demonstrate the individual intensity profiles focussed on the MFP peak only, showing electric field-induced peak shift indicating a variation of the MFP phase lattice parameter(s). This observation unequivocally proves the piezoelectricity of the MFP phase. As expected, the bulk lattice parameters remain fixed for the centrosymmetric $Pm\bar{3}m$ STO. It also does not show any sign of electrostriction, which would be allowed for the centrosymmetric space group. We must stress that this piezoelectric response has dynamical nature. While permanent removal of the electric field promptly destroys the MFP phase ([11]) its cyclic (1 kHz) variation modifies its lattice parameters only.

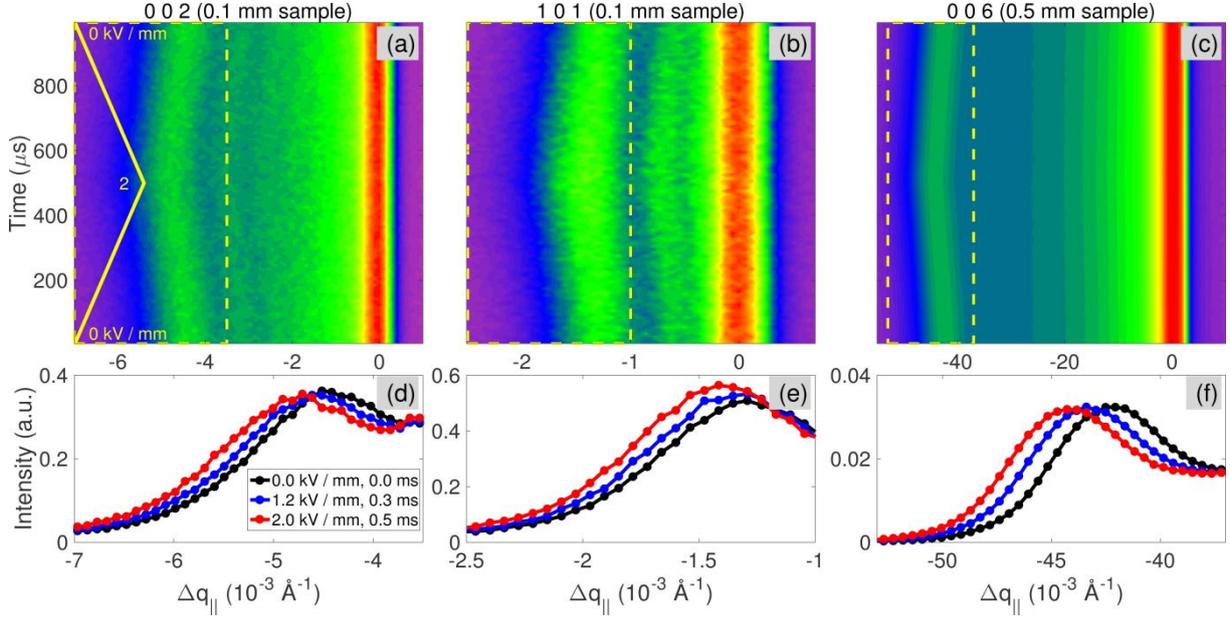

**Figure 3.** Time–resolved X-ray diffraction scans through the Bragg reflections under cyclic electric field. The average electric field is equal to 1 kV/mm, while the time-dependent 1 kHz electric field varies between 0 kV/mm and 2 kV/mm. (a-c) are false-colour maps of diffraction intensity of 002 and 101 and 006 reflections. (d-f) display individual line profiles, clearly demonstrating displacements of the MFP peak position due to the variation of electric field.

We quantified the newly discovered piezoelectric activity by fitting one-dimensional intensity profiles with the sum of two pseudo-Voigt functions (using FITYK package [32]). To reduce the noise, we binned the original 10 000 time channels into 50 by summing up the intensities of each 200 adjacent ones. The bulk peak was fitted using only the points on the right of its maximum, whereas the MFP peak was fitted using only the points on the left of its maximum (Figs. 4(a-b)). The resulting electric field dependencies of the MFP peak position in Figs. 4(c-d) show a linear response of the corresponding lattice spacing to the electric field in all cases. Because of the tetragonal symmetry of the MFP phase, the piezoelectric response to the electric field, applied along [001] direction, is described by two independent piezoelectric coefficients $d_{33} = \frac{1}{c}\frac{\partial c}{\partial E}$ and $d_{31} = \frac{1}{a}\frac{\partial a}{\partial E}$. Taking into account that $q_{\parallel[00l]} = \frac{2\pi l}{c}$ and $q_{\parallel[101]} = 2\pi\sqrt{\frac{1}{a^2}+\frac{1}{c^2}}$, we can express these coefficients as:

$$d_{33} = \frac{-1}{q_{\parallel[00l]}}\frac{\partial(q_{\parallel[00l]})}{\partial E} \qquad (1)$$

$$d_{31} = \frac{-a^2}{c^2}d_{33} - a^2 q_{\parallel[101]}\frac{\partial(q_{\parallel[101]})}{\partial E} \qquad (2)$$

Subsequently, we obtained for the 0.1 mm thick sample $d_{33} = (64.3 \pm 0.6)$ pC/N and $d_{31} = (-2.8 \pm 1.2)$ pC/N as well as $d_{33} = (107.4 \pm 0.6)$ pC/N for the 0.5 mm thick sample. Hence, a fivefold increase of the sample thickness increased the piezoelectric coefficient by a factor of 1.7.

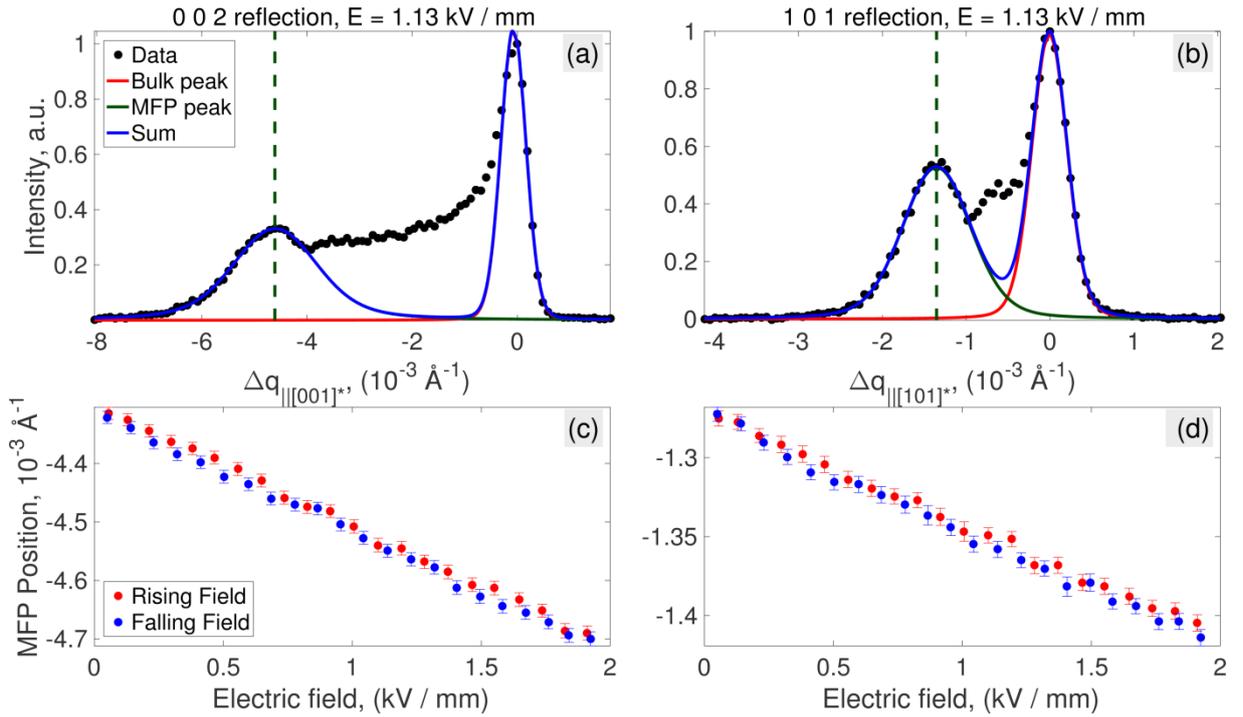

**Figure 4.** Decomposition of the measured radial intensity profiles into the components, corresponding to bulk STO reflection and MFP phase reflection: (a) and (b) show the measured and fitted diffraction curves at one selected field amplitude of $E = 1.13$ kV/mm for 002 and 101 reflections, respectively. The variation of the MFP phase peak positions (vertical lines in (a) and (b)) as a function of applied electric field is shown in (c-d) for the 002 and 101 reflection correspondingly.

In summary, this letter reports the evidence of artificially produced piezoelectricity in the near-surface layer of originally centrosymmetric STO single crystals. This layer can be created[11] by exposing STO to a static electric field for a few hours, redistributing the oxygen vacancies towards one side of the plate. Because of the interlinked STO bulk, the properties of the MFP phase cannot be easily accessed via macroscopic measurements. On the contrary, *in-situ* X-ray diffraction enabled us to separate and simultaneously follow the field- and time-dependence of lattice parameters in both, the active MFP phase and the passive STO bulk. Importantly, the piezoelectricity of the MFP phase is relatively high: It can compete with the piezoelectricity in perovskite-based functional materials (e.g. PZT [21]).

However, a high piezoelectric response in ferroelectric materials is often ascribed to the presence of ferroelastic domains and the motion of domain walls between them. At the same time, for the symmetry reasons, the MFP phase can only be formed in a tetragonal single domain state. Its piezoelectricity is therefore clearly intrinsic. Finally, the thicker sample exhibited a higher amount of strain as well as an increased value for $d_{33}$. As the amount of migration-induced field-stabilized strain in STO depends on many factors like real structure or the concentration of oxygen vacancies, the functional dependencies between thickness, strain and piezoelectric coefficients are not immediately obvious and merit future investigation. We may also expect that e.g. creating natural barriers for migration of oxygen vacancies inside the STO bulk could shift the piezoelectric activity from the surface in the bulk.

**Acknowledgments**

This work was supported by German Federal Ministry of Education and Research (C. Richter: BMBF 05K13OF1; E. Mehner, J. Hanzig, M. Zschornak, T. Leisegang 03EK3029A; B.

Khanbabaee, S. Gorfman, 05K13PSA). The authors would like to thank ESRF and beamline BM28-XMaS staff for their support.

**References**

[1]   J. Schooley, W. Hosler, and M. L. Cohen, Physical Review Letters **12**, 474 (1964).
[2]   J. Robertson, Journal of Vacuum Science & Technology B **18**, 1785 (2000).
[3]   A. Sawa, Materials today **11**, 28 (2008).
[4]   J. Mannhart and D. Schlom, Science **327**, 1607 (2010).
[5]   M. G. Walter, E. L. Warren, J. R. McKone, S. W. Boettcher, Q. Mi, E. A. Santori, and N. S. Lewis, Chemical reviews **110**, 6446 (2010).
[6]   R. Waser, R. Dittmann, G. Staikov, and K. Szot, Advanced Materials **21**, 2632 (2009).
[7]   N. H. Chan, R. Sharma, and D. M. Smyth, Journal of The Electrochemical Society **128**, 1762 (1981).
[8]   K. Szot, W. Speier, R. Carius, U. Zastrow, and W. Beyer, Physical Review Letters **88**, 075508 (2002).
[9]   T. Leisegang, H. Stöcker, A. Levin, T. Weißbach, M. Zschornak, E. Gutmann, K. Rickers, S. Gemming, and D. Meyer, Physical Review Letters **102**, 087601 (2009).
[10]  H. Stöcker, M. Zschornak, J. Seibt, F. Hanzig, S. Wintz, B. Abendroth, J. Kortus, and D. C. Meyer, Applied Physics A **100**, 437 (2010).
[11]  J. Hanzig *et al.*, Physical Review B **88**, 024104 (2013).
[12]  Y. Li, Y. Lei, B. Shen, and J. Sun, Scientific reports **5** (2015).
[13]  C. Ang, Z. Yu, and L. Cross, Physical Review B **62**, 228 (2000).
[14]  K. Szot, W. Speier, G. Bihlmayer, and R. Waser, Nature materials **5**, 312 (2006).
[15]  J. Blanc and D. L. Staebler, Physical Review B **4**, 3548 (1971).
[16]  J. Hanzig, M. Zschornak, M. Nentwich, F. Hanzig, S. Gemming, T. Leisegang, and D. C. Meyer, Journal of Power Sources **267**, 700 (2014).
[17]  Y. Lei *et al.*, Nature communications **5** (2014).
[18]  J. Hanzig, E. Mehner, S. Jachalke, F. Hanzig, M. Zschornak, C. Richter, T. Leisegang, H. Stöcker, and D. C. Meyer, New Journal of Physics **17**, 023036 (2015).
[19]  W. Heywang, K. Lubitz, and W. Wersing, *Piezoelectricity: evolution and future of a technology* (Springer Science & Business Media, 2008), Vol. 114.
[20]  R. Bechmann, Physical review **110**, 1060 (1958).
[21]  G. H. Haertling, Journal of the American Ceramic Society **82**, 797 (1999).
[22]  S. Brown, P. Thompson, M. Cooper, J. Kervin, D. Paul, W. Stirling, and A. Stunault, Nuclear Instruments and Methods in Physics Research Section A: Accelerators, Spectrometers, Detectors and Associated Equipment **467**, 727 (2001).
[23]  R. J. Harrison, S. A. Redfern, A. Buckley, and E. K. Salje, Journal of Applied Physics **95**, 1706 (2004).
[24]  J. Leist, H. Gibhardt, K. Hradil, and G. Eckold, Journal of Physics: Condensed Matter **23**, 305901 (2011).
[25]  G. Eckold, H. Schober, and S. E. Nagler, *Studying kinetics with neutrons* (Springer, 2010).
[26]  J. Y. Jo, P. Chen, R. J. Sichel, S. J. Callori, J. Sinsheimer, E. M. Dufresne, M. Dawber, and P. G. Evans, Physical Review Letters **107**, 055501 (2011).

[27]    S. Gorfman, O. Schmidt, M. Ziolkowski, M. von Kozierowski, and U. Pietsch, Journal of Applied Physics **108**, 064911 (2010).
[28]    S. Gorfman, O. Schmidt, V. Tsirelson, M. Ziolkowski, and U. Pietsch, Zeitschrift für anorganische und allgemeine Chemie **639**, 1953 (2013).
[29]    S. Gorfman, Crystallography Reviews **20**, 210 (2014).
[30]    S. Gorfman, H. Simons, T. Iamsasri, S. Prasertpalichat, D. Cann, H. Choe, U. Pietsch, Y. Watier, and J. Jones, Scientific reports **6** (2016).
[31]    S. Gorfman, H. Choe, V. V. Shvartsman, M. Ziolkowski, M. Vogt, J. Strempfer, T. Łukasiewicz, U. Pietsch, and J. Dec, Physical Review Letters **114**, 097601 (2015).
[32]    M. Wojdyr, Journal of Applied Crystallography **43**, 1126 (2010).